\documentclass[]{spie}  
\usepackage[]{graphicx}

\title{Fifteen years of the Advanced CCD Imaging Spectrometer} 

\author{Catherine E. Grant\supit{a}, Mark W. Bautz\supit{a}, Peter G. Ford\supit{a} and Paul. P. Plucinsky\supit{b}
\skiplinehalf
\supit{a}Kavli Institute for Astrophysics and Space Research, Massachusetts Institute of Technology, Cambridge, Massachusetts, USA; \\
\supit{b}Harvard-Smithsonian Center for Astrophysics, Cambridge, Massachusetts, USA
}

\authorinfo{Further author information: (Send correspondence to C.E.G.)\\C.E.G.: E-mail: cgrant@space.mit.edu}

\begin{document} 
  \maketitle 

\begin{abstract}
As the Advanced CCD Imaging Spectrometer (ACIS) on the Chandra X-ray Observatory enters its fifteenth year of operation on orbit, it continues to perform well and produce spectacular scientific results. The response of ACIS has evolved over the lifetime of the observatory due to radiation damage, molecular contamination and aging of the spacecraft in general. Here we present highlights from the instrument team's monitoring program and our expectations for the future of ACIS. The ACIS calibration source produces multiple line energies and fully illuminates the entire focal plane which has greatly facilitated the measurement of charge transfer inefficiency and absorption from contamination. While the radioactive decay of the source has decreased its utility, it continues to provide valuable data on the health of the instrument. Performance changes on ACIS continue to be manageable, and do not indicate any limitations on ACIS lifetime.
\end{abstract}


\section{INTRODUCTION}
\label{sec:intro}

The Chandra X-ray Observatory, the third of NASA's great observatories in space, was launched just past midnight on July 23, 1999, aboard the space shuttle {\it Columbia}\cite{cha2}.  After a series of orbital maneuvers, Chandra reached its operational orbit, with initial 10,000-km perigee altitude, 140,000-km apogee altitude, and 28.5$^\circ$ inclination.  In this evolving high elliptical orbit, Chandra transits a wide range of particle environments, from the radiation belts at closest approach through the magnetosphere and magnetopause and past the bow shock into the solar wind.

The Advanced CCD Imaging Spectrometer (ACIS), one of two focal plane science instruments on Chandra, utilizes charge-coupled devices (CCDs) of two types, front- and back-illuminated (FI and BI).  Soon after launch it was discovered that the FI CCDs had suffered radiation damage from exposure to soft protons scattered off the Observatory's grazing-incidence optics during passages through the Earth's radiation belts\cite{gyp00}.  The BI CCDs were unaffected.  Since mid-September 1999, ACIS has been protected during radiation belt passages and there is an ongoing effort to prevent further damage and to develop hardware and software strategies to mitigate the effects of charge transfer inefficiency on data analysis. 

As ACIS enters its fifteenth year of operation on orbit, it continues to perform well and produce spectacular scientific results.  The response of ACIS has evolved over the lifetime of the observatory due to radiation damage, molecular contamination and degradation of thermal control.  The ACIS instrument team has an ongoing monitoring program, has developed software and calibration products to improve instrument response and flight software patches that further protect the instrument to help ensure a long lifetime.  We discuss some of these efforts here.

\section{DATA}

Most of the results discussed use data taken of the ACIS External Calibration Source (ECS) which fully illuminates the ACIS CCDs while ACIS is stowed.  Since the discovery of the initial radiation damage, a continuing series of observations of the ECS have been undertaken just before and after the instruments are safed for perigee passages to monitor the performance of the ACIS CCDs.  ACIS is placed in the stowed position exposing the CCDs to the ECS which produces many spectral features, the strongest of which are Mn-K$\alpha$ (5.9~keV), Ti-K$\alpha$ (4.5~keV), and Al-K (1.5~keV).  The data are taken in the standard Timed Exposure mode with a 3.2~second frame time.  Typical exposure times for each observation range from 5.5 to 8~ksecs.  All standard observations are taken at the nominal focal plane temperature set-point of --120$^\circ$~C.

\begin{figure}
\begin{center}
\includegraphics[height=3.5in]{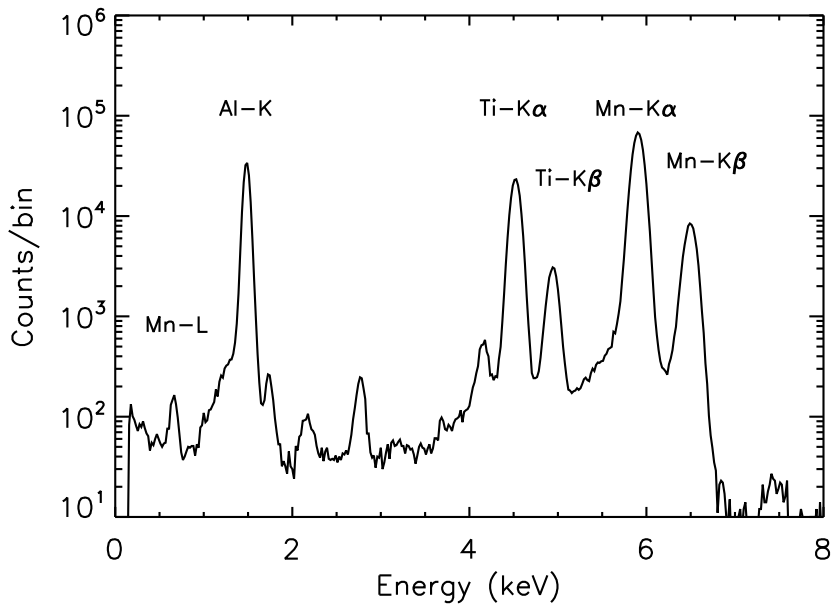}
\caption{The spectrum of the ACIS External Calibration Source on the ACIS-I3 FI CCD.  Only data from the first 256 rows are included to minimize CTI degradation.  Lines used for calibration and monitoring purposes are labelled.  The remaining features are instrumental.}
\label{fig:spec}
\end{center}
\end{figure}

The Fe-55 radioactive source produces strong lines of Mn-K$\alpha$ and Mn-K$\beta$ at 5.9~keV and 6.5~keV and a much weaker complex of unresolved Mn-L lines at $\sim$700 eV.  Fluorescence targets add lines from Al-K at 1.5~keV and Ti-K$\alpha$ and Ti-K$\beta$ at 4.5~keV and 4.9~keV.  The access to multiple energies has proven invaluable in studying gain changes on short time-scales and the energy dependence of charge transfer inefficiency, as well as the change in low-energy efficiency due to molecular contamination.  Figure~\ref{fig:spec} shows the spectrum of the ACIS External Calibration Source and labels the important features.

The frequent cadence of ECS observations, roughly twice every three days, allows for close monitoring of ongoing radiation damage and rapid investigation of potential instrument anomalies, while still providing a long-baseline for detailed studies of performance changes.  In addition to the regular ECS observations near perigee, much longer exposures are often undertaken during long duration radiation shutdowns due to solar storms while the external environment is still too active to restart science observations.

The full illumination of the focal plane by the ECS enables mapping of detector performance down to the level of a few pixels.  In particular, the charge transfer inefficiency correction included in the standard pipeline processing utilizes maps of the electron trap uniformity.\cite{cticorr}  These maps include the known gradient in the radiation damage across the focal plane, as well as variations in charge loss from column to column due to the variation in the number of traps in each column.  A substantial fraction of the improvement provided by the CTI correction is due to the mapping of this fine structure in the trap density.

ACIS uses a combination of passive radiators and active heaters to maintain thermal control.  While the focal plane temperature set point has remained at --120$^\circ$~C for most of the Chandra mission, thermal control has become more difficult with time, particularly for some spacecraft orientations and near perigee where the solid angle of the warm Earth as viewed by the ACIS radiator can be significant.  Because they occur close to perigee, ECS observations are much more likely than science observations to be warmer than nominal by as much as 5--10$^\circ$~C in the worst cases.  Depending on the type of monitoring or calibration activity and how much temperature dependence is expected, analysis of the warmer data may just combine it with cold data, may apply a temperature-correction first and then combine, or it may be disregarded for that particular calibration activity.  The availability of warmer calibration data has proved invaluable for developing a temperature-dependent charge transfer inefficiency correction that is now part of standard pipeline processing.\cite{ctitemp}.

\begin{figure}
\begin{center}
\includegraphics[height=3.5in]{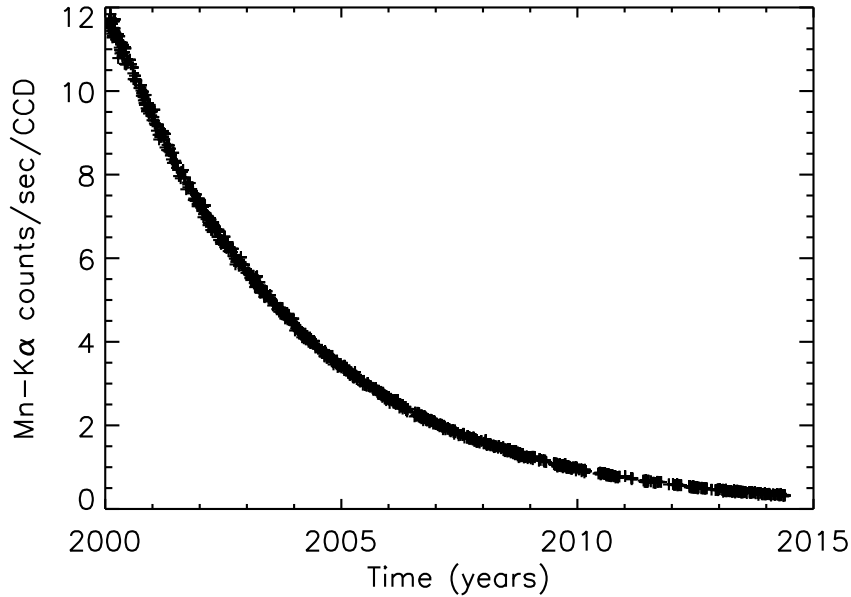}
\caption{The decline in the count rate of the ACIS ECS due to radioactive decay of the Fe-55 source.}
\label{fig:kdecay}
\end{center}
\end{figure}

The ECS is powered by the radioactive isotope Fe-55 which has a half-life of 2.737 years. Figure~\ref{fig:kdecay} shows the decline in the count rate of the Mn-K$\alpha$ line throughout the Chandra mission due to the decay of the radioactive source. Clearly over the fifteen year lifetime of Chandra, the count rate from the ECS has dropped substantially requiring more thoughtful analysis and some reduction in the ability to simultaneously measure spatial and temporal effects.  The weakest line from the low energy Mn-L complex at $\sim$700 eV is essentially lost in the underlying background spectrum after roughly 2010.  The stronger lines remain well measured and should continue to provide valuable calibration and monitoring data into the future.  

\section{CHARGE TRANSFER INEFFICIENCY}

Our primary measure of radiation damage on the CCDs is charge transfer inefficiency (CTI).  The eight front-illuminated CCDs had essentially no CTI before launch, but are strongly sensitive to radiation damage from low energy protons ($\sim$100~keV) which preferentially create traps in the buried transfer channel.  The framestore covers, 2.5-mm of gold coated aluminum, are thick enough to stop this radiation, so the initial damage was limited to the imaging area of the FI CCDs.  Radiation damage from low-energy protons is now minimized by moving the ACIS detector away from the aimpoint of the observatory during passages through the Earth's particle belts and during solar storms.  Continuing exposure to both low and high energy particles over the lifetime of the mission slowly degrades the CTI further.\cite{odell,ctitrend}  The two back-illuminated CCDs (ACIS-S1,S3) suffered damage during the manufacturing process and exhibit CTI in both the imaging and framestore areas and the serial transfer array.  However, owing to their much deeper charge-transfer channel, BI CCDs are insensitive to damage by the low-energy protons that damage FI devices.

\begin{figure}
\centering
\includegraphics[width=3.3in]{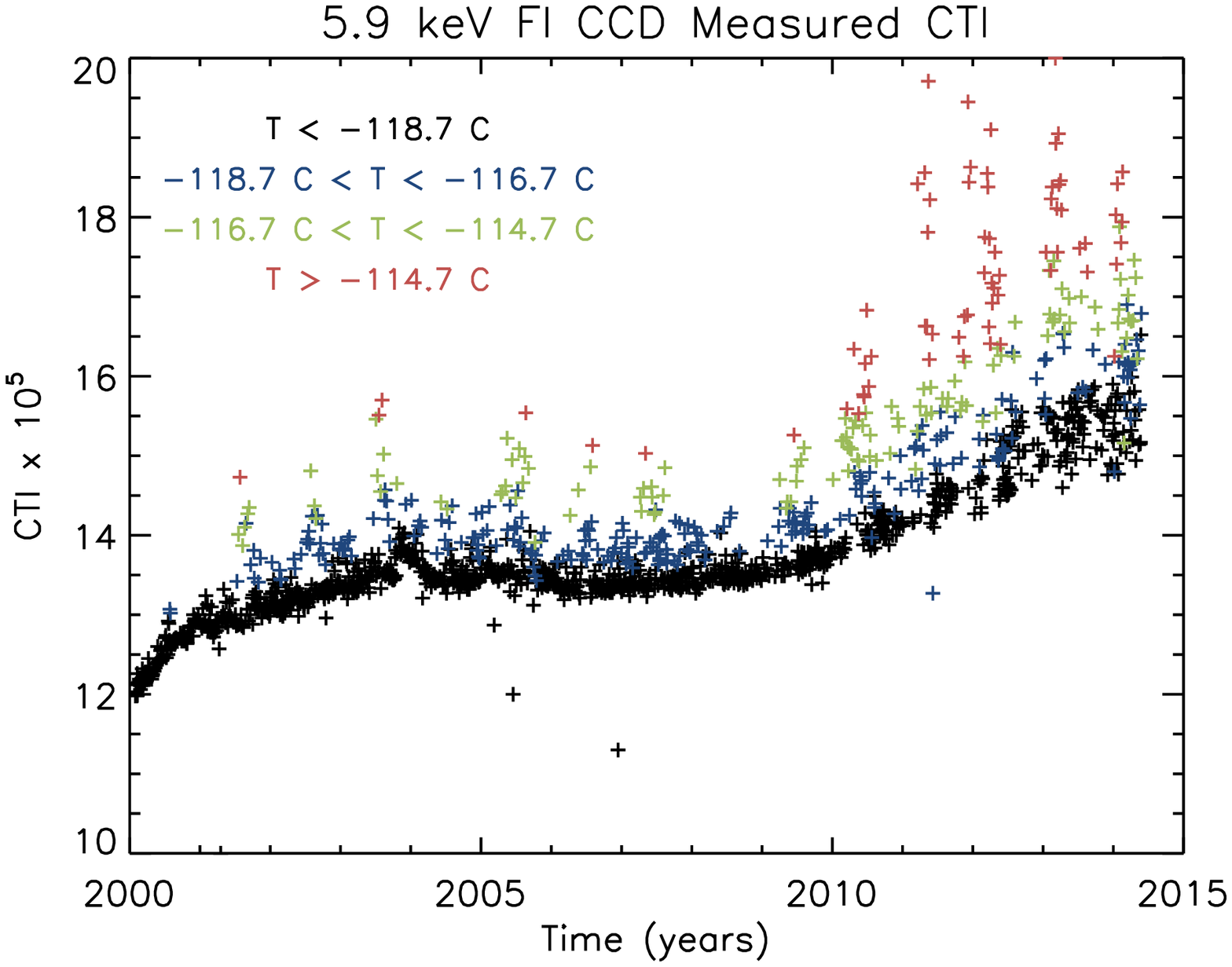}%
\hfill%
\includegraphics[width=3.3in]{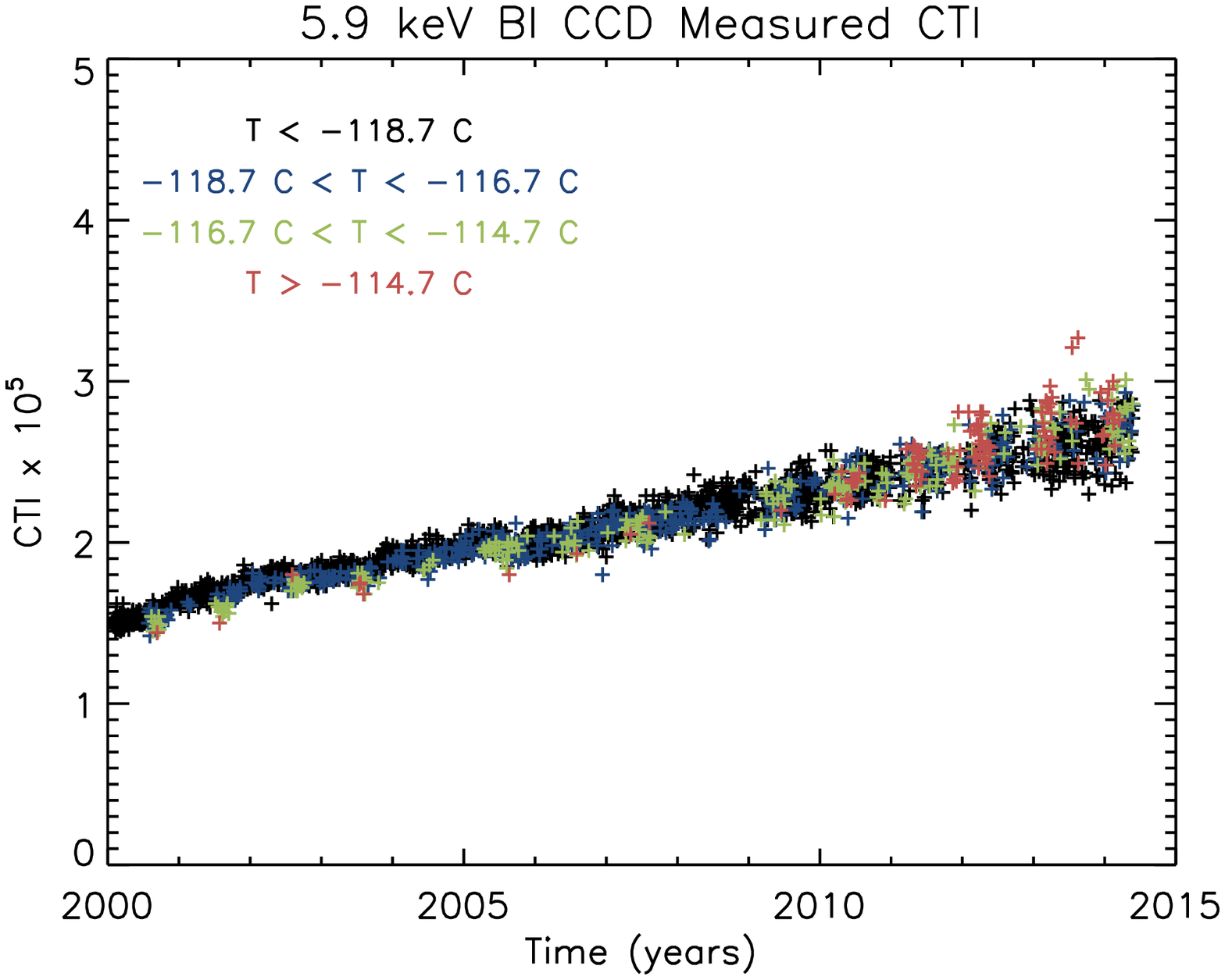}
\caption{Evolution of measured CTI for the FI (left) and BI (right) CCDs.  Each point is a single ECS observation.  The colors indicate times of warmer focal plane temperature which increases CTI.  The remaining structure is due to accumulated radiation damage modulated by sacrificial charge from the changing particle background.  The BI CCD has lower CTI with much less dependence on temperature and particle background than the FI CCD.} \label{fig:ctiuncorr}
\end{figure}

Figure~\ref{fig:ctiuncorr} shows the measured parallel CTI from January 2000 through May 2014 without any corrections for temperature or sacrificial charge from the particle background.  The mean of the four ACIS-I CCDs (I0-I3) are used to typify the behavior of the FI CCDs, while the ACIS-S3 CCD is used for BI CCDs.  The FI CCDs have much higher CTI, due to the initial radiation damage from low energy protons.  The type of electron traps in the FI CCDs are much more sensitive to variations in temperature which cause the upward scatter in the CTI data, and to sacrificial charge from the particle background, which causes the sawtooth shapes in the lower data points.  The temperature dependence of the initial BI CTI is the reverse of the FI CCDs, with higher temperatures causing lower measured CTI.  As the mission continues and the BI CCDs accumulate radiation damage, the temperature dependence is changing to be more like the FI CCDs.

\begin{figure}
\centering
\includegraphics[height=3.5in]{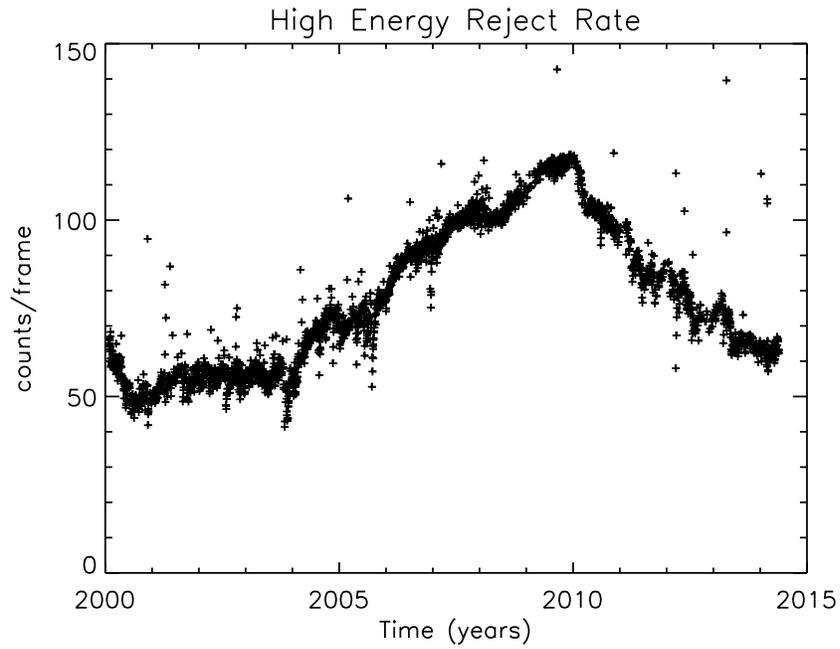}
\caption{Time dependence of the particle background as measured by the rate of high energy events on ACIS-S3 (BI).  The longest scale variation is due to the solar cycle and the smaller features to individual solar storms.} \label{fig:bkg}
\end{figure}

The variation in the charged particle background is shown in figure~\ref{fig:bkg}.  We use the rate of events with energies too high to be X-rays focused by the telescope as a proxy for the particle rate.  These events have been shown to be well correlated with $> 10$~MeV protons and are anti-correlated with the solar cycle.\cite{bkg}  Additional smaller scale dips and transient increases are due to solar storms.  These features in the particle background can be seen in reverse to some extent in the measured FI CTI in Figure~\ref{fig:ctiuncorr} due to sacrificial charge.

\begin{figure}
\centering
\includegraphics[width=3.3in]{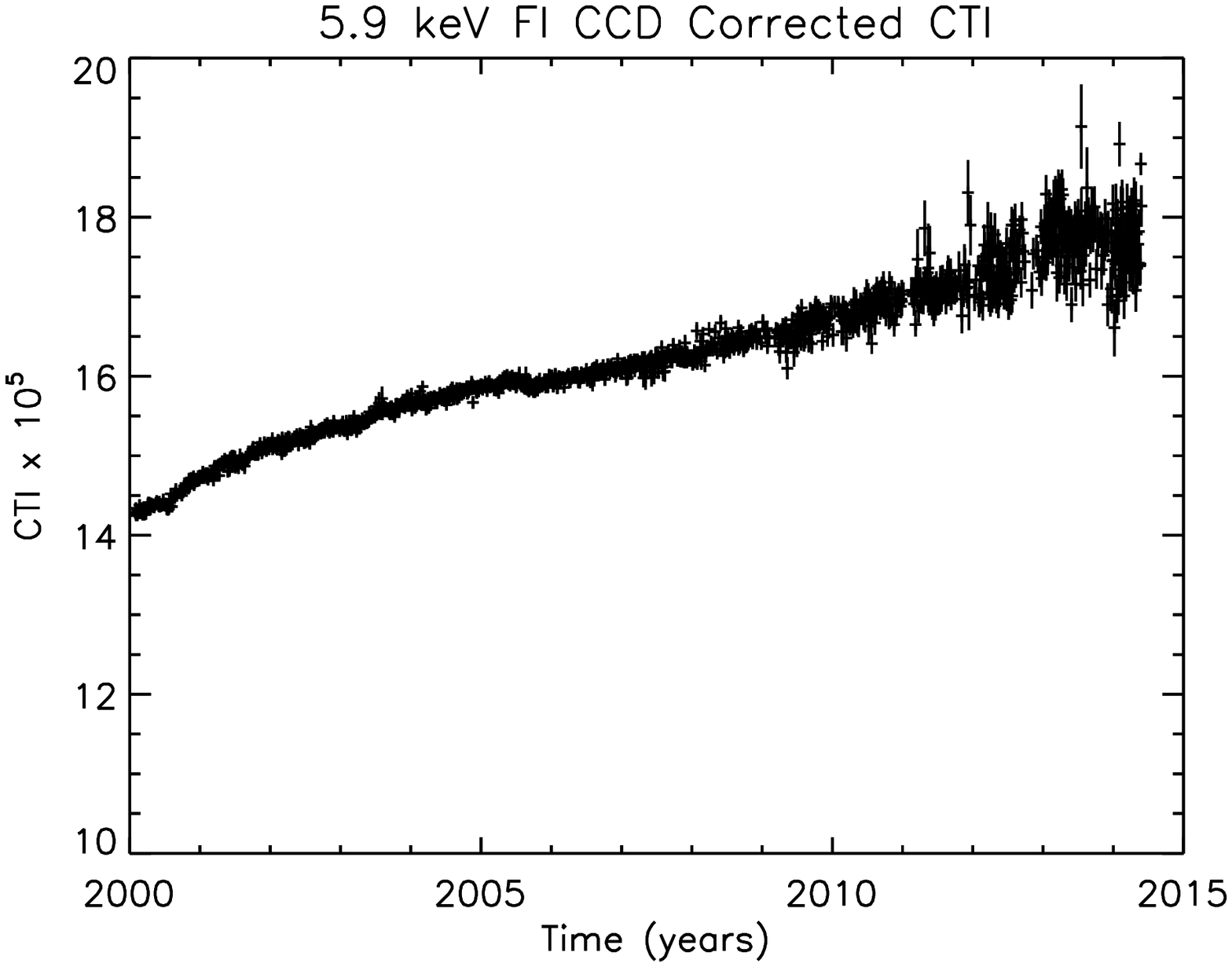}%
\hfill%
\includegraphics[width=3.3in]{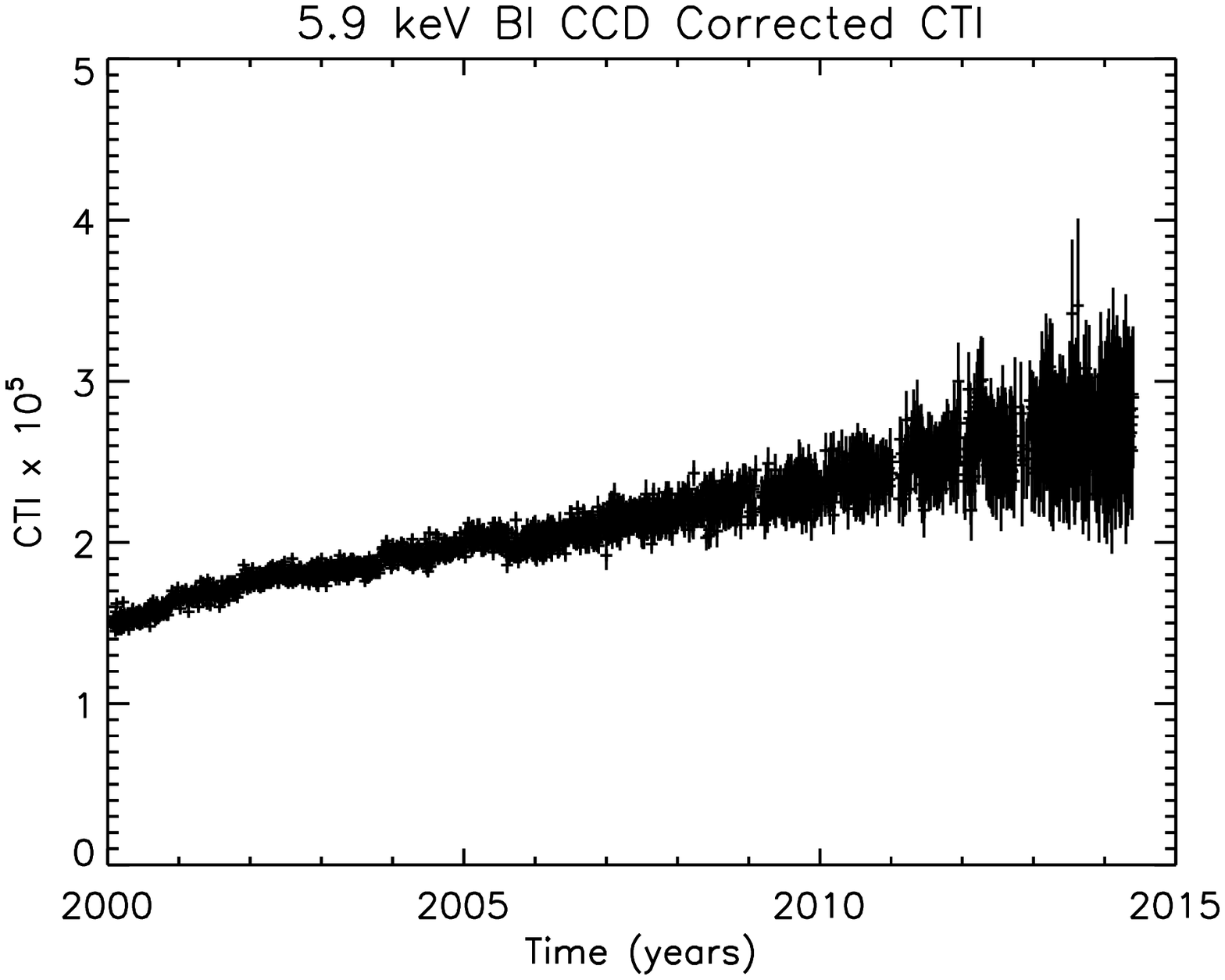}
\caption{Evolution of corrected CTI for the FI (left) and BI (right) CCDs. These data have had corrections applied which remove variations due to temperature and sacrificial charge from the particle background. } \label{fig:cticorr}
\end{figure}

Figure~\ref{fig:cticorr} shows the same CTI data as Figure~\ref{fig:ctiuncorr} after applying corrections for temperature variations and sacrificial charge from the particle background.  The scatter is lower, but is increasing toward the end of the time period due to the reduced count rate from the decaying radioactive source.  The CTI increase appears slow and smooth, with no evidence for step function changes due to specific solar events.  The Chandra radiation management plan appears to be effective at limiting ACIS exposure to damaging low energy protons.\cite{odell}  The CTI increase of the FI CCDs is higher than that of the BI CCD which indicates that the accumulated radiation damage must include a mixture of particle energies, including the low energy protons which preferentially damage the FI CCDs but not the BI CCDs.

Table~\ref{tab:ctiincrease} lists the initial CTI and the rate of increase after the corrections for temperature and particle background.  The table also includes measurements and upper limits on serial CTI as well as CTI in the framestore array.  The measurements of CTI in the framestore array are indirect, as we can only measure the pulseheight change at the bottom of the imaging array, rather than the full pulseheight versus position fit, but a small change in framestore CTI is possibly seen for both types of CCDs.  Measuring serial CTI on the FI CCDs is difficult because of the strong parallel CTI component, but the FI CCDs do not yet have any measurable serial CTI.  The serial CTI of the BI CCD was measurable at launch and is increasing very slowly.  The framestore and serial transfer arrays are shielded by the framestore cover which reduces their exposure to radiation damage.

\begin{table}
\vspace{0.09in}
\caption{CTI Increase Summary}
\label{tab:ctiincrease}
\begin{center}
\begin{tabular}{lccc}
&Initial CTI (2000) &Yearly Increase  &Yearly Increase\\
&($10^{-5}$) &($10^{-6}$ / yr) &( \% / yr) \\
\hline \\
FI Parallel CTI    &$14.32 \pm 0.04$ &$2.29 \pm 0.01$ &$1.60 \pm 0.01$ \\
FI Framestore CTI$^*$ &$\cdots$         &$0.60 \pm 0.02$ &$\cdots$ \\
FI Serial CTI     &$< 0.7$            &$< 0.89$ &$\cdots$ \\
\\
BI Parallel CTI   &$1.51 \pm 0.03$  &$0.83 \pm 0.01$ &$5.50 \pm 0.02$ \\
BI Framestore CTI$^*$ &$\cdots$         &$0.94 \pm 0.05$   &$\cdots$ \\
BI Serial CTI     &$8.27 \pm 0.43$  &$0.22 \pm 0.04$ &$0.26 \pm 0.05$ \\

\end{tabular}

\smallskip

\noindent
\small
$^*$ Assuming no change in electronic gain. \\
\normalsize
\end{center}
\end{table}

\section{RADIATION MONITORING}

The Chandra team has implemented procedures to protect ACIS during high-radiation events including autonomous protection triggered by an on-board radiation monitor. Elevated temperatures have reduced the effectiveness of this on-board monitor. The ACIS team has developed an algorithm which uses data from the CCDs themselves to detect periods of high radiation\cite{radmon1} and a flight software patch to apply this algorithm is currently active on-board the instrument.\cite{radmon2,pgfradmon} 

Since early in the Chandra mission, procedures have been implemented that protect the focal plane instruments during times of high radiation. \cite{odell} ACIS is translated out of the focal plane, providing protection against soft protons, and is powered off.  Three types of procedures are in place; planned protection during radiation-belt transits, autonomous protection triggered by the on-board radiation monitor, and manual intervention based upon assessment of space-weather conditions.  The Chandra weekly command load includes automatic scheduled safing of the focal-plane instruments during radiation belt passages.  The timings of radiation belt ingress and egress are determined using the standard AP8/AE8 environment with a small additional pad time to protect against temporal variations.  Solar storms are detected either by the on-board radiation monitor or by ground operations monitoring of various space weather measures, such as from NASA's Advanced Composition Explorer (ACE), the NOAA Geostationary Operational Environmental Satellite system (GOES), and the planetary Kp index.  The on-board radiation monitor cannot detect protons at hundreds of keV, which are the most damaging to ACIS, so on-board protection is supplemented by other measures of the radiation environment. 

Due to elevated temperatures, the on-board radiation monitor has become increasingly unreliable and as of November 2013 is no longer part of the autonomous radiation protection plan.  Concern about the effectiveness of the on-board particle monitor motivated exploring other on-board measures of the radiation environment such as ACIS itself.\cite{radmon1}

The flight software patch was installed on ACIS in November 2011 and updated with more optimal parameters in April 2012.   The Chandra On-Board Computer was patched in May 2012 and will now respond to any ACIS radiation triggers with the standard radiation protection procedures.  The patch has operated as expected with no impact on regular science operations and has correctly detected an enhanced radiation environment on two occasions and triggered autonomous radiation protection of the spacecraft. To better match the changing quiescent particle background which is anti-correlated with the solar cycle, the parameters of the patch will be re-evaluated a few times a year and updated as necessary.

\begin{figure}
\begin{center}
\includegraphics[height=3.5in]{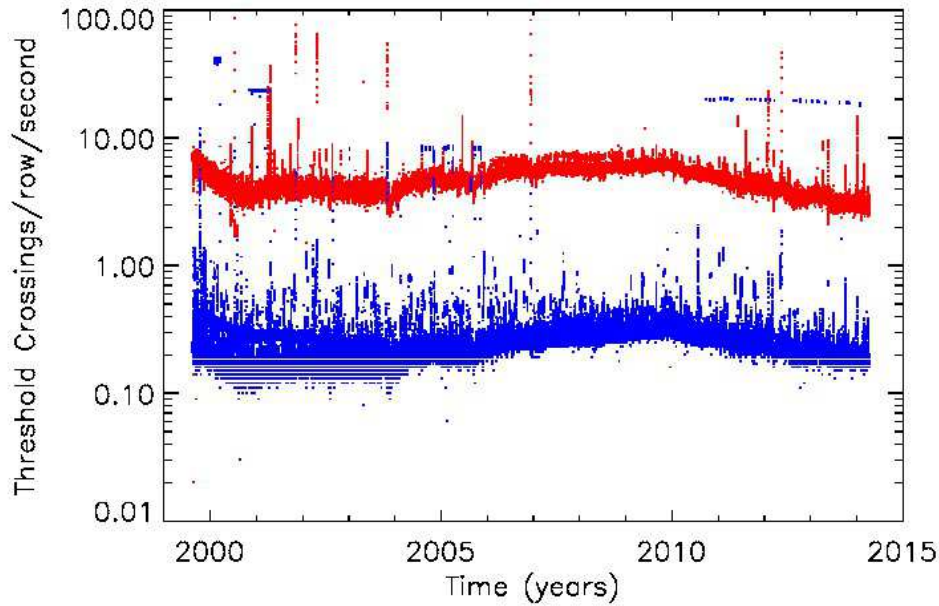}
\caption{ACIS threshold crossing rate as a function of time over the entire history of Chandra.  The (red) upper points denote the mean of the FI CCDs; the (blue) lower points, the BI CCDs.  The features are explained in the text.}
\label{fig:threshplot}
\end{center}
\end{figure}

The ACIS flight software patch uses the event threshold crossing rate to monitor the radiation environment.  In particular, it keeps track of the threshold crossings per second per row, which is, for the most part, independent of the instrument observing mode.  To reduce the noise in the threshold rates, the radiation monitor patch examines the mean of the active FI CCDs and the mean of the active BI CCDs, in three minute time bins.  Figure~\ref{fig:threshplot} shows the ACIS threshold crossing rate as a function of time from August 1999 through May 2012.  These data include contributions from astrophysical X-rays and particles.

A number of features can be seen in Figure~\ref{fig:threshplot}.  The threshold crossing rate of the FI CCDs is about an order of magnitude higher than that of the BI CCDs due to the structural differences of the two types of CCDs.  Particle interactions in the FI CCDs produce large blooms due to the thicker active and field-free regions.  The BI CCDs have a thinner active region and no field-free region, so the particle events occupy much smaller areas and produce fewer pixels above the event threshold.  There is a long timescale structure in the quiescent rates with lower threshold crossing rates during 2001--2004 and 2013--present, with higher rates in 2008--2010, due to the 11-year solar cycle.  During solar maximum the sun's magnetic field provides extra shielding against cosmic rays which depresses the threshold crossing rate.

Times of enhanced threshold crossing rates are also seen in Figure~\ref{fig:threshplot}. The vertical enhancements correspond to real increases in the particle environment associated with solar activity.  The horizontal features seen in the BI CCD data are due to repeated observations of a specific bright X-ray source, the Crab Nebula.  The radiation detection algorithm used by ACIS differentiates between high count rates due to bright sources and radiation events by examining the rise time of the light curve.

\section{MOLECULAR CONTAMINATION}

\begin{figure}
\begin{center}
\includegraphics[height=3.5in]{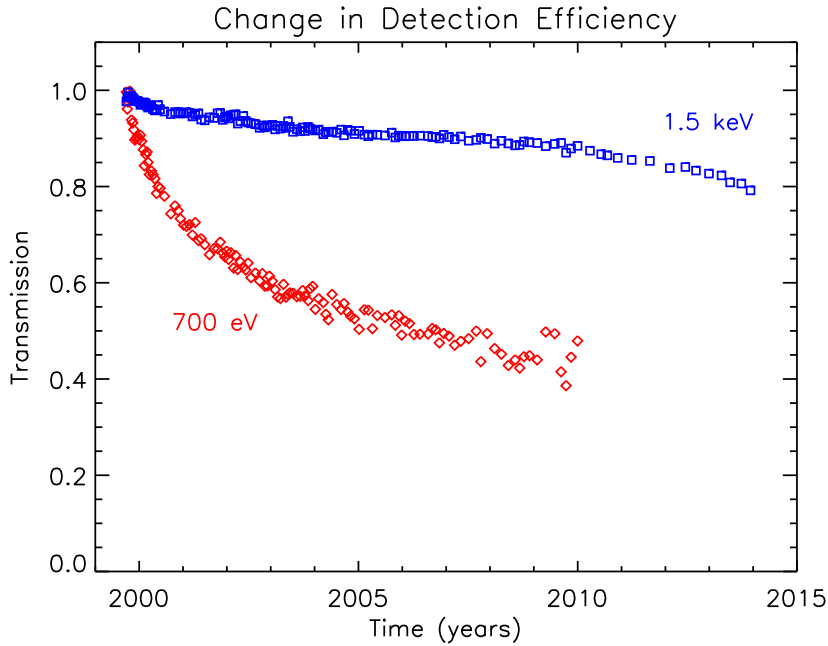}
\caption{The slow decrease in low energy X-ray transmission due to accumulating contamination on the optical blocking filter.  These data are from the ACIS-S3 CCD summed over the entire active area.  The 700~eV data have very low signal-to-noise after 2010 and are not plotted.}
\label{fig:contam}
\end{center}
\end{figure}

The low energy efficiency of ACIS has been impacted by the slow accumulation of a molecular contaminant on the optical blocking filter.\cite{contamHLM,contamSOD}  The decrease in efficiency due to the contaminant can be monitored using the ACIS ECS; in particular by comparing the ratio of low energy lines such as Al-K at 1.5~keV and the Mn-L complex at $\sim$700~eV, to that of Mn-K$\alpha$ at 5.9~keV.  The full illumination of the focal plane by the ECS has facilitated the spatial mapping of the contaminant, which is thicker near the edges of the focal plane than in the middle.  Figure~\ref{fig:contam} shows the change in low energy efficiency due to increasing contamination at 700~eV and 1.5~keV.  Due to the decay of the radioactive source and the intrinsic faintness of the 700~eV line complex, the lowest energy data become unreliable after 2010 and are not plotted.  After the initial fast increase in contamination deposition, the rate appeared to slow down until roughly 2010 when the rate increased again.  The reason for the rate of increase in recent times is currently under study, but may be related to increasing spacecraft temperatures.  

\section{FUTURE PROSPECTS}

The lifetime of Chandra is not currently limited by consumables or orbital changes and there are no known limitations to a mission of twenty-five years or longer.\cite{chandra15} Gradual degradation of thermal shielding has required some changes in mission planning and operations which reduces flexibility, but does not pose a hazard to continued operation.  Anomalies in ACIS have been infrequent and with limited impact on science.  Most instrument anomalies have been cleared by the power cycling that is part of the standard autonomous commands between every observation.  

Performance changes on ACIS continue to be manageable, and do not indicate any limitations on ACIS lifetime.  At the current rate of CTI increase, changes in gain and spectral resolution are small, requiring periodic changes to calibration, but not impacting potential science results.  At some locations, far from the framestore where the effects of CTI are strongest, the pulseheights of the very lowest energy events ($< 400$~eV) can approach the event detection threshold, and may not be detected, but the number of impacted events are small and should not increase dramatically.  For the FI CCDs, with much larger initial CTI, the additional accumulated CTI is small and represents only a minor modification.  The BI CCDs with lower initial CTI have larger relative performance changes, but the absolute level remains manageable.

The loss of low energy efficiency due to contamination is potentially more significant.  The reason behind the time evolution is not understood so extrapolating calibration measurements to future observations can be less accurate than desirable and leads to a delay between the time of an observation and the time when calibration products correctly capture the performance at that time.  Observations for which low energy X-rays are desired will require increasingly longer exposure times to reach the same data quality.  The ACIS instrument team and CXC calibration groups are actively working toward improving our knowledge of the ACIS contamination and continuing to pursue mitigation strategies.

As maintaining the nominal focal plane temperature becomes more difficult, it may be determined that raising the temperature set point is necessary.  This would increase the CTI and require a much more substantial recalibration effort, as well as decrease the spectral resolution.\cite{ctitemp}.  Until that time, software tools have been developed which can follow small temperature variations during an observation and adjust the calibration products accordingly.

Finally, due to radioactive decay the utility of the ACIS External Calibration Source will continue to decrease.  Observations of astrophysical sources will have to replace the functionality of the ECS observations. Some calibration tasks, such as monitoring the thickness of the contamination layer, have already moved to primarily using astrophysical targets.  Other tasks, such as monitoring CTI, have continued to use the ECS, although with poorer statistics than early in the mission.  The ACIS calibration team is currently developing a plan for a smooth transition from the ECS to astrophysical sources for all calibration tasks, while still preserving as much long-term calibration accuracy as possible.

We look forward to at least another fifteen years of spectacular ACIS science results.

\acknowledgments

We are exceedingly grateful to all the scientists and engineers involved in building and supporting the continued operation of the Chandra X-ray Observatory.  In particular, we would like to thank the Chanda Science Operations team, and the Chandra Project Science team for many years of fruitful collaboration and constant vigilance.  This work was supported by NASA contracts NAS 8-37716 and NAS 8-38252. 

\bibliography{article} 
\bibliographystyle{spiebib} 

\end{document}